\documentclass[
  journal=pasa,
  manuscript=research-paper, 
  year=2022,
  volume=,
]{cup-journal}
\pdfoutput=1
\usepackage{microtype,siunitx,booktabs}
\usepackage{float}
\usepackage{amsmath}	
\usepackage{amssymb}	
\usepackage{multicol}   
\usepackage{bm}		    
\usepackage{pdflscape}	
\usepackage{array}
\usepackage{booktabs}
\usepackage{tabularx}
\usepackage{natbib}
\usepackage{rotating}
\usepackage{subfigure}
\usepackage{diagbox}
\usepackage{fancyhdr}
\usepackage{longtable}
\usepackage{latexsym}
\usepackage{graphicx}
\usepackage{amsmath}
\usepackage{natbib}
\usepackage{upgreek}
\usepackage{multirow}
\usepackage{array}
\usepackage{pdflscape}
\usepackage{graphicx,subfigure}
\usepackage{tabularx}
\usepackage{epsfig}
\usepackage{color}
\usepackage{lscape}
\usepackage{ulem}
\usepackage{algorithmicx,algorithm}
\usepackage{diagbox}
\usepackage{makecell}
\usepackage{hyperref}
\usepackage{enumitem}
\usepackage{tablefootnote}
\usepackage{threeparttable}
\usepackage[final]{changes}
\usepackage{array}

\definecolor{myblue}{RGB}{0, 0, 255} 
\setaddedmarkup{\textcolor{myblue}{#1}} %

\defcitealias{ho2020}{Ho20}

\title{Photometric redshift estimation for emission line galaxies of DESI Legacy Imaging Surveys by CNN-MLP}

\author{Shirui Wei}
\affiliation{National Astronomical Observatories, Chinese Academy of Sciences, Beijing 100012, China.}
\alsoaffiliation{University of Chinese Academy of Sciences, Beijing, 100049, China.}
\alsoaffiliation{National Astronomical Data Center, Beijing 100101, China.}

\author{Changhua Li}
\affiliation{National Astronomical Observatories, Chinese Academy of Sciences, Beijing 100012, China.}
\alsoaffiliation{National Astronomical Data Center, Beijing 100101, China.}

\author{Yanxia Zhang}
\affiliation{National Astronomical Observatories, Chinese Academy of Sciences, Beijing 100012, China.}

\author{Chenzhou Cui}
\affiliation{National Astronomical Observatories, Chinese Academy of Sciences, Beijing 100012, China.}
\alsoaffiliation{National Astronomical Data Center, Beijing 100101, China.}

\author{Chao Tang}
\affiliation{National Astronomical Observatories, Chinese Academy of Sciences, Beijing 100012, China.}
\alsoaffiliation{University of Chinese Academy of Sciences, Beijing, 100049, China.}
\alsoaffiliation{National Astronomical Data Center, Beijing 100101, China.}

\author{Jingyi Zhang}
\affiliation{National Astronomical Observatories, Chinese Academy of Sciences, Beijing 100012, China.}

\author{Yongheng Zhao}
\affiliation{National Astronomical Observatories, Chinese Academy of Sciences, Beijing 100012, China.}

\author{Xue-Bing Wu}
\affiliation{Department of Astronomy, School of Physics, Peking University, Beijing, 100871, China.}
\alsoaffiliation{Kavli Institute for Astronomy and Astrophysics, Peking University, Beijing 100871, China.}

\author{Yihan Tao}
\affiliation{National Astronomical Observatories, Chinese Academy of Sciences, Beijing 100012, China.}
\alsoaffiliation{National Astronomical Data Center, Beijing 100101, China.}

\author{Dongwei Fan}
\affiliation{National Astronomical Observatories, Chinese Academy of Sciences, Beijing 100012, China.}
\alsoaffiliation{National Astronomical Data Center, Beijing 100101, China.}

\author{Shanshan Li}
\affiliation{National Astronomical Observatories, Chinese Academy of Sciences, Beijing 100012, China.}
\alsoaffiliation{National Astronomical Data Center, Beijing 100101, China.}

\author{Yunfei Xu}
\affiliation{National Astronomical Observatories, Chinese Academy of Sciences, Beijing 100012, China.}
\alsoaffiliation{National Astronomical Data Center, Beijing 100101, China.}

\author{Maoyuan Huang}
\affiliation{National Astronomical Observatories, Chinese Academy of Sciences, Beijing 100012, China.}
\alsoaffiliation{National Astronomical Data Center, Beijing 100101, China.}

\author{Xingyu Yang}
\affiliation{National Astronomical Observatories, Chinese Academy of Sciences, Beijing 100012, China.}
\alsoaffiliation{National Astronomical Data Center, Beijing 100101, China.}

\author{Zihan Kang}
\affiliation{National Astronomical Observatories, Chinese Academy of Sciences, Beijing 100012, China.}
\alsoaffiliation{University of Chinese Academy of Sciences, Beijing, 100049, China.}

\author{Jinghang Shi}
\affiliation{National Astronomical Observatories, Chinese Academy of Sciences, Beijing 100012, China.}
\alsoaffiliation{University of Chinese Academy of Sciences, Beijing, 100049, China.}

\email[Changhua Li; Yanxia Zhang; Chenzhou Cui]{lich@bao.ac.cn; zyx@bao.ac.cn; ccz@bao.ac.cn}


\begin{document}

\begin{abstract}
Emission Line Galaxies (ELGs) are crucial for cosmological studies, particularly in understanding the large-scale structure of the Universe and the role of dark energy. ELGs form an essential component of the target catalogue for the Dark Energy Spectroscopic Instrument (DESI), a major astronomical survey. However, the accurate selection of ELGs for such surveys is challenging due to the inherent uncertainties in determining their redshifts with photometric data. In order to improve the accuracy of photometric redshift estimation for ELGs, we propose a novel  approach CNN--MLP that combines Convolutional Neural Networks (CNNs) with Multilayer Perceptrons (MLPs). This approach integrates both images and photometric data derived from the DESI Legacy Imaging Surveys Data Release 10. By leveraging the complementary strengths of CNNs (for image data processing) and MLPs (for photometric feature integration), the CNN--MLP model achieves a $\sigma_{\mathrm{NMAD}}$ (normalised median absolute deviation) of 0.0140 and an outlier fraction of 2.57\%. Compared to other models, CNN--MLP demonstrates a significant improvement in the accuracy of ELG photometric redshift estimation, which directly benefits the target selection process for DESI. In addition, we explore the photometric redshifts of different galaxy types (Starforming, Starburst, AGN, Broadline). Furthermore, this approach will contribute to more reliable photometric redshift estimation in ongoing and future large-scale sky surveys (e.g. LSST, CSST, Euclid), enhancing the overall efficiency of cosmological research and galaxy surveys.
\end{abstract}
\section{Introduction} \label{sec:intro}
Redshifts play an important role in studying galaxy evolution and cosmology. Based on measurement methods, redshifts can be categorised into two types: spectroscopic redshift ($z_\mathrm{{spec}}$) and photometric redshift ($z_\mathrm{{phot}}$). $z_\mathrm{{spec}}$ is determined by leveraging multiple characteristic features of spectra, such as emission and absorption lines, and calculating the stretch amount of their wavelengths. With high spectral resolution, for example $R = \lambda / \Delta \lambda$ > 200, $z_\mathrm{{spec}}$ can offer superior accuracy with precision better than $10^{-3}$ \citep{Salvato2019}. However, obtaining spectra is time-consuming, leading to a substantial fraction of sources without associated spectra to compute redshifts. In contrast, $z_\mathrm{{phot}}$ measurement relies on fluxes from a set of filters, meaning that we can derive redshifts for all sources identified in a multi-band imaging survey, making it a popular option for estimating redshifts in current and future imaging surveys despite its lower precision, 
e.g. the Dark Energy Survey \citep[DES;][]{DES2016}, the Euclid survey \citep{Euclid2024} , the Vera C. Rubin Observatory Legacy Survey of Space and Time \citep[LSST;][]{LSST2019} and the China Space Station Telescope \citep[CSST;][]{CSST}. 

Since the idea of $z_\mathrm{{phot}}$ first implemented by \cite{Baum1962}, many methods have been proposed to advance the estimation precision. The modern mainstream methods can be concluded into two primary approaches: template fitting and machine learning. Template fitting methods estimate $z_\mathrm{{phot}}$ by comparing the observed spectral energy distributions (SEDs) with a set of reference spectra, namely templates and $z_{\mathrm{phot}}$ is determined by the best fit, like LePhare \citep{Laphare1,Laphare2}, BPZ \citep{BPZ}, HYPERZ \citep{HYPERZ}, ZEBRA \citep{ZEBRA} and EAZY \citep{EAZY}. Machine learning, on the other hand, predicts redshifts by learning a mapping relationship between photometry and $z_\mathrm{{spec}}$ based on vast representative training samples. Classic machine learning algorithms used in photometric redshift estimation include Support Vector Machines \citep[SVM;][]{SVM}, $k$ nearest neighbours \citep[kNN;][]{kNN1,kNN2}, Random Forest \citep[RF;][]{Randomforest}, XGBoost \citep{XGBoost} and Catboost \citep{Li24}, etc.

Each approach has its own merits and drawbacks. In a recent study by \cite{Li24}, they selected two representative methods from both template fitting and machine learning approaches, EAZY and CatBoost respectively, to study the performance of $z_\mathrm{{phot}}$ using the Dark Energy Spectroscopic Instrument Legacy Imaging Surveys Data Release 10 \citep[DESI LS10;][]{DESI2019} catalogue. The results demonstrated that machine learning methods perform better when trained on a comprehensive dataset that covers a sufficiently wide parameter space. While for some faint and high-redshift samples which are underrepresented in the training set, machine learning methods are constrained by the generalisation issue. In contrast, template-fitting method can make predictions over a broader redshift range, as it does not rely on training samples with a fixed redshift range. Nevertheless, its performance depends on the completeness of the templates and the assumptions on galaxy physics \citep{Dey2022}, making it prone to producing outliers when confronted with novel galaxy spectral types or unmatched templates.

The Emission Line Galaxies (ELGs), following the spectral classification of  the Sloan Digital Sky Survey (SDSS; \citealt{York2000}), include starforming, starburst, AGN and broadline galaxies. They contain rich information about the processes occurring within galaxies, including star formation, ionisation, and chemical evolution. ELGs remain a major focus of DESI experiment \citep{DESI2013,DESI2016a,DESI2016b}, comprising one-third of the survey targets. They serve as key tracers for probing the large-scale structure over the redshift range $0.6 < z < 1.6$, and provide the tightest DESI cosmological constraints \citep{ELGtarget}. The primary ELG target selection is based on two key criteria: a magnitude cut in the $g$-band and a colour selection box defined in ($g - r$) versus ($r - z$) space. The complete selection details can be found in Table 2 of \cite{ELGtarget}. Accurate estimation of $z_\mathrm{{phot}}$ plays a crucial role in the pre-selection of ELG targets, minimising the fraction of non-ELG sources while simultaneously improving the completeness of the selected objects. By integrating multi-band photometric data, estimating $z_\mathrm{phot}$ avoids the need for direct emission-line detection. Additionally, it allows for cross-validation with existing $z_\mathrm{spec}$ measurements, which enhances the confidence in target selection, reduces spectroscopic incompleteness, and ultimately supports the scientific goals of the DESI mission \citep{Newman2022}. 

However, compared with the Bright Galaxy Sample (BGS) and Luminous Red Galaxies (LRGs), also major targets in DESI, $z_\mathrm{{phot}}$ estimation for ELGs remains significantly more challenging and less accurate. For instance,  \cite{Zhou24_DESIEDR} reported that the catastrophic outlier rate (i.e., the fraction of objects whose $z_\mathrm{{phot}}$ deviate significantly from their $z_\mathrm{{spec}}$)  reaches 15.78\% for ELGs in the DESI Early Data Release \citep[EDR;][]{DESIEDR} survey validation programme \citep{ELGtarget}. In contrast, the corresponding outlier rates for BGS and LRGs are only  0.83\% and 1.07\%, respectively. Similar results have also been reported in \cite{Zhou2023ELG} and \cite{Li24}.
Compared with other types of galaxies, the ELGs are usually fainter in $z$-band, with higher redshifts. Moreover, the correlation between photometric features (e.g., from DESI $g,r,z$ and unWISE $W1,W2$ bands, as well as half-light radius) and redshift is substantially weaker, resulting in lower accuracy of photometric redshifts of ELGs \citep{Zhou24_DESIEDR}.
These findings highlight the importance of developing improved feature extraction methods for faint, high-redshift ELG sources, and of identifying photometric features that are more strongly correlated with redshift for more accurate estimation.

To enhance the performance of $z_\mathrm{{phot}}$ estimation, recent advances in deep neural networks, particularly convolutional neural networks \citep[CNN;][]{CNN}, have facilitated significant progress due to their ability to extract image-based features effectively. Unlike photometric data, imaging data offer more intuitive insights, including morphology, apparent size, and surface brightness. Consequently, several studies have explored the use of multi-band galaxy images as inputs for redshift prediction, instead of relying solely on manually extracted photometric features \citep{Hoyle16,Pasquet19,Dey2022}. Given the complementary nature of photometric and imaging data, a number of studies have tried to integrate these two data types. This approach leverages the unique strengths of each type of data, offering a more comprehensive understanding of galaxy properties, resolving degeneracies, and achieving superior results compared to methods that use a single data type \citep{Henghes22,Yao23,Zhang24,Roster24}. Inspired by previous work, we put forward a CNN-MLP model to improve $z_\mathrm{{phot}}$ performance for ELGs. 

The paper is organised as follows. In Section 2, we describe the ELG sample and the corresponding images and photometric data utilised in this study. In Section 3, we introduce the model architecture that we propose to combine multi-band images and photometric features. Section 4 outlines the evaluation metrics and experimental setup we employ. In Section 5, we present and analyse the $z_\mathrm{{phot}}$ estimation results. Finally, Section 6 provides a summary of our findings and conclusions.

\section{The Data} \label{sec:data}
\subsection{Photometric data} \label{Photometric_Data}
The DESI Legacy Imaging Surveys (DESI LS) provide galaxy and quasar targets for follow-up observation by DESI. The original Legacy Surveys comprise three individual surveys: the Dark Energy Camera (DECam) Legacy Survey \citep[DECaLS;][]{DECam,DESI2019}, the Beijing-Arizona Sky Survey \citep[BASS;][]{BASS} and the Mayall z-band Legacy Survey \citep[MzLS;][]{MzLS}. Together, these surveys cover 14000 $\mathrm{deg^2}$ in three optical bands ($g$, $r$, $z$), additionally augmented with four infrared bands ($W1$, $W2$, $W3$, $W4$) from the Near-Earth Object Wide-field Infrared Survey Explorer Reactivation Mission \citep[NEOWISE;][]{NEOWISE}. The DESI LS10 is the tenth public data release of the DESI LS and it comprises two distinct parts, the north of the galactic plane imaged by BASS/MzLS and the south of the Galactic plane imaged by DECam. Notably, the data from the southern sky of DESI LS10 include not only $g$, $r$, $z$-band observations from DECaLS, but also $g$, $r$, $i$, $z$-band DECam observations from several non-DECaLS surveys, i.e., the Dark Energy Survey \citep[DES;][]{DES2016}, the DECam Local Volume Exploration Survey \citep[DELVE;][]{DELVE} and the DECam eROSITA Survey (DeROSITAs, PI: A. Zenteno, Zenteno et al. in prep). These observations expand the coverage to more than 20000 $\mathrm{deg^2}$. The images and photometric data of the ELG sources used in this study are from the south of the galactic plane imaged by DECam in DESI LS10. 

The catalogue of DESI LS10 provides aperture and model fluxes as well as the morphological model types used in the Tractor fitting procedure. We select galaxies classified as REX (round exponential galaxies with a variable radius), DEV (deVaucouleurs elliptical galaxies), EXP (exponential spiral galaxies) and SER (Sersic profiles). The AB magnitudes corresponding to the model and aperture fluxes are then computed for the $g$, $r$, $i$, $z$, $W1$ and $W2$ bands. Sources exceeding the magnitude limits of $g$ > 24.0, $r$ > 23.4, $z$ > 22.5 are excluded, aligned with DESI requirements. Additionally, sources with maskbits != 0 are removed, as the flag indicates issues such as pixel saturation or contamination from nearby bright stars, globular clusters, or large galaxies. To ensure data integrity, we also exclude sources with missing flux values in any bands. Following these criteria, the photometric data of galaxies are obtained.

\subsection{Spectroscopic redshifts} \label{sec:spec_redshift}
The spectroscopic redshifts are used as the true value labels to train a supervised photometric redshift estimation model. The ELGs with known spectroscopic redshifts are mainly composed of three parts. The first part is from the SDSS DR17 spectroscopic catalogue \citep{SDSSDR17}, selecting sources with CLASS == GALAXY and SUBCLASS as one of STARFORMING, STARBURST, AGN and BROADLINE.
The second part is from the ELG targets of DESI EDR \citep{DESIEDR} sv1 and sv3. The third part is from different spectroscopic surveys which are cross-matched with SIMBAD within a 1 arc second radius, and those sources whose main type is AGN or EmG are kept. These surveys 
include the Two Degree Field Galaxy Redshift Survey \citep[2dFGRS;][]{2dFGRS}, the Six Degree Field Galaxy Survey \citep[6dFGS;][]{6dFGS}, the Complete Calibration of the Color-Redshift Relation survey \citep[C3R2;][]{C3R2}, the VIMOS VLT Deep Survey \citep[VVDS;][]{VVDS}, zCOSMOS \citep{zCOSMOS}, the DEEP2 Galaxy Redshift Survey \citep[DEEP2;][]{DEEP2}, the PRIsm MUlti-object Survey \citep[PRIMUS;][]{PRIMUS1,PRIMUS2}, the Fiber Multi-Object Spectrograph COSMOS survey \citep[FCOSMOS;][]{FCOSMOS}, the VIMOS Public Extragalactic Redshift Survey \citep[VIPERS;][]{VIPERS}, the Large Sky Area Multi-Object Fiber Spectroscopic Telescope \citep[LAMOST;][]{LAMOST1,LAMOST2,LAMOST3} DR8, the WiggleZ Dark Energy Survey \citep[WiggleZ;][]{WiggleZ}, the Galaxy And Mass Assembly \citep[GAMA;][]{GAMA} DR3 and Optical redshifts for the Dark Energy Survey \citep[OzDES;][]{OzDES}.

Among these ELGs, spectroscopic measurements with low-quality flags or redshifts less than or equal to zero are excluded based on the quality criteria provided in the respective catalogues. Thus the ELG sources with known spectroscopic redshifts are obtained.

\subsection{Known ELG sample} \label{Emission Line Galaxy Sample}
\begin{figure}[ht]
    \centering
    \includegraphics[width=0.9\textwidth]{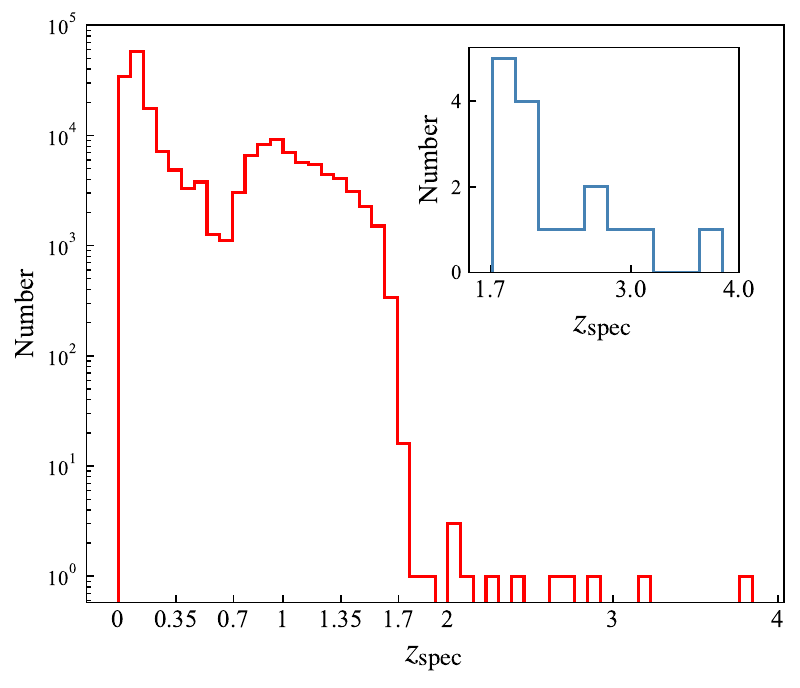}
    \caption{$z_\mathrm{{spec}}$ distribution for the ELG sample. The majority of sources have $z_\mathrm{{spec}}$ values below 1.7, with two prominent peaks in the low-redshift region. The inset panel in the upper right corner displays the distribution of the relatively rare high-redshift sources with $z_\mathrm{spec}$ greater than 1.7.}
    \label{fig:redshift_distribution}
\end{figure}

\begin{figure}[ht]
    \centering
    \includegraphics[width=0.9\textwidth]{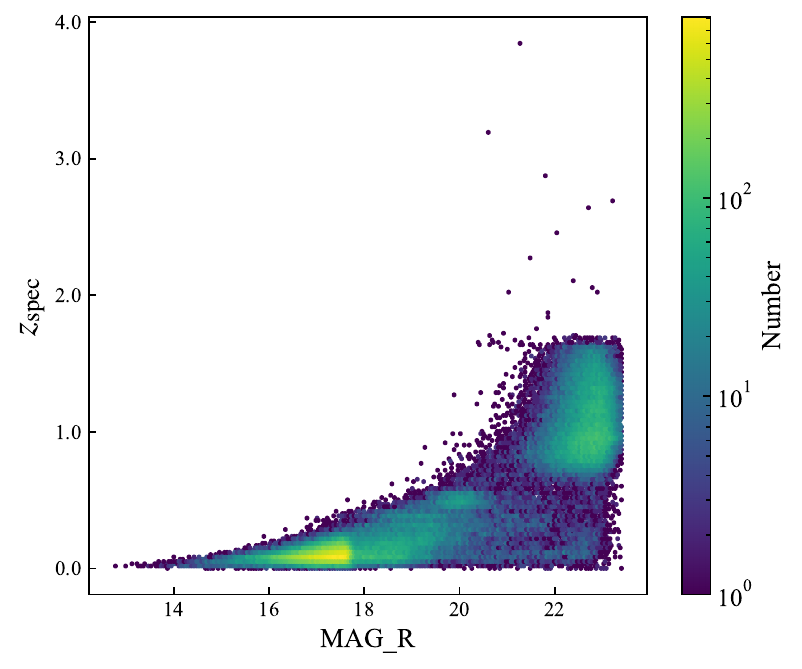}
    \caption{$z_\mathrm{{spec}}$ v.s. $r$-band magnitude. A clear trend is evident, with low-redshift sources generally being brighter and high-redshift sources tending to be fainter.}
    \label{fig:redshift_r}
\end{figure}

The ELG sources with spectroscopic redshifts are cross-matched with  the galaxy photometric data described in Section \ref{Photometric_Data} within a 1 arc second radius. The nearest matches within this radius are retained, forming the known ELG sample for training and testing. The ELG sample comprises 192375 unique entries. Details of the spectroscopic surveys and their selection criteria included in the ELG sample are provided in Table \ref{tab:survey_info}. Figure~\ref{fig:redshift_distribution} illustrates the distribution of $z_\mathrm{{spec}}$, and Figure~\ref{fig:redshift_r} describes the relationship between $z_\mathrm{{spec}}$ and $r$-band magnitude within this sample. As illustrated in Figure~\ref{fig:redshift_distribution}, the $z_\mathrm{spec}$ values in the sample range from 0 to 4, with the majority falling below 1.7. The overall redshift distribution exhibits two prominent peaks, separated approximately at $z \sim 0.7$. The first peak, centred around $z \sim 0.1$, corresponds primarily to sources from the SDSS survey, while the second peak, located near $z \sim 1.0$, mainly reflects contributions from the DESI EDR. This bimodal distribution is consistent with the sample statistics summarized in Table \ref{tab:survey_info}.

Additionally, Figure~\ref{fig:redshift_r} shows that high-redshift galaxies tend to be faint, with two noticeable clusters in the $r$-band magnitude ranges of 16-18 and 22-23.

\begin{table*}[ht]
\centering
\caption{Information about the various spectroscopic surveys included in the dataset. $\Tilde{z}_{spec}$ denotes the median of $z_\mathrm{{spec}}$. $\Tilde{r}$(mag) denotes the median of $r$-band magnitude.}
\begin{tabular}{ccccc}
\toprule
Survey    & No. of sources     & Quality Criterion & $\Tilde{z}_{\rm spec}$ & $\Tilde{r}$(mag) \\ 
\hline
SDSS      & 122452  &    zWarning=0               & 0.11         & 17.48 \\
DESI SV3  & 52491   &   zWarning=0, ZERR<0.01                & 1.04 & 22.73 \\
DESI SV1  & 10782   &   zWarning=0, ZERR<0.01                & 0.94 & 22.63 \\
6dFGRS    & 3220    &    Q=3,4               & 0.06          & 15.57 \\
PRIMUS    & 909     &    ZQUALITY=3,4               & 0.56          & 21.37 \\
zCOSMOS   & 716     &     5$\textgreater$CC$\geq$3              & 0.68             & 21.67 \\
LAMOST    & 612     &  Z\_ERR$\leq$ 0.01                & 0.12 & 16.98 \\
GAMA      & 442     &   NQ=4                & 0.19             & 18.78 \\
VIPERS    & 274     &  zflag=3,4                 & 0.79 & 21.83 \\
WiggleZ   & 118     &  qz=4,5                 & 0.63           & 20.90 \\
VVDS      & 109     &   ZFLAGS=4                & 0.74             & 22.13 \\
fCOSMOS   & 103     &   ZFLAG $\geq$ 3                & 1.48 & 22.73 \\
2dFGRS    & 88      &    Q=4               & 0.11             & 17.38 \\
OzDES     & 28      &   qop=4                & 0.46 & 20.46 \\
C3R2      & 19      &   Qual$\geq$3                & 0.82              & 22.43 \\
DEEP2     & 12      &   ZQUALITY$\geq$3                & 0.82          & 22.67 \\
\hline
Total No. & 192375 &  & 0.16 & 18.41 \\
\hline
\end{tabular}

\label{tab:survey_info}
\end{table*}

All relevant parameters of the known ELG sample are summarised in Table \ref{tab:catalog_param}, including aperture magnitudes, model magnitudes and galactic extinction values E(B-V). 
For the imaging data, we utilise the FITS file cutout service provided by DESI LS\footnote{\url{https://www.legacysurvey.org/viewer/fits-cutout}} to download the corresponding images in the $g$, $r$, $i$, $z$, $W1$ and $W2$ bands, based on the RA and DEC coordinates of a source. Although the projections of image stacks around the brick centre are identical in both optical and infrared bands, ensuring consistent coordinates across different bands, the pixel resolution differs. The pixel scale is 0.262 arc seconds per pixel for $g$, $r$, $i$, $z$ bands and 2.75 arc seconds per pixel for $W1$ and $W2$ bands. These default pixel scales are maintained to ensure consistency with source detection and photometry configurations. The image size is chosen to be 64 $\times$ 64 pixels, as this size is sufficient to capture the background sky and the full extent of each galaxy while minimising contamination from other objects. Each source is centred in the image through the cutout service. Finally, the resulting image data obtained from the service have a shape of (192375, 6, 64, 64), with the bands arranged in the order: $g$, $r$, $i$, $z$, $W1$ and $W2$.

\begin{table*}[ht]
\centering
\begin{threeparttable}
\caption{Photometric data corresponding to the known ELG dataset.}
\begin{tabular}{llll}
\toprule
Name    & Definition                 & Catalogue & Waveband \\ 
\midrule
MAG\_G            & Model magnitude in $g$ band & DESI          & Optical band   \\
MAG\_R            & Model magnitude in $r$ band & DESI          & Optical band   \\
MAG\_I            & Model magnitude in $i$ band & DESI          & Optical band   \\
MAG\_Z            & Model magnitude in $z$ band & DESI          & Optical band   \\
MAG\_W1           & Model magnitude in $W1$ band & DESI         & Infrared band  \\
MAG\_W2           & Model magnitude in $W2$ band & DESI         & Infrared band  \\
APMAG\_G\_(1$\sim$8)\tnote{a}    & Aperture magnitude in $g$ band & DESI       & Optical band   \\
APMAG\_R\_(1$\sim$8)     & Aperture magnitude in $r$ band & DESI       & Optical band   \\
APMAG\_I\_(1$\sim$8)     & Aperture magnitude in $i$ band & DESI       & Optical band   \\
APMAG\_Z\_(1$\sim$8)     & Aperture magnitude in $z$ band & DESI       & Optical band   \\
APMAG\_W1\_(1$\sim$5)\tnote{b}    & Aperture magnitude in $W1$ band & DESI      & Infrared band  \\
APMAG\_W2\_(1$\sim$5)    & Aperture magnitude in $W2$ band & DESI      & Infrared band  \\
ebv & Galactic extinction E(B-V) reddening  & DESI \\ 
z\_spec      & Spectroscopic redshift    & Known sample  &    \\ 
\bottomrule
\end{tabular}
\begin{tablenotes}
\footnotesize
\item[a] Optical aperture magnitude suffixes 1$\sim$8 correspond to fluxes measured within circular apertures of radii [0.5, 0.75, 1.0, 1.5, 2.0, 3.5, 5.0, 7.0] arcsec.
\item[b] Infrared aperture magnitude suffixes 1$\sim$5 correspond to apertures with radii [3, 5, 7, 9, 11] arcsec.
\end{tablenotes}
\label{tab:catalog_param}
\end{threeparttable}
\end{table*}

\subsection{Feature construction}
Features are an important factor that influences model performance. For photometric data,  we combine all model magnitudes with aperture magnitudes listed in Table~ \ref{tab:catalog_param} and compute various differences between them, forming 97 features in total. From these, we select 85 features to construct the optimal feature set that yields the best performance. Details of the feature selection procedure and the complete list of selected features is provided in \ref{sec:feature_selection}. For imaging data, we combine the raw images with colour images, resulting in a channel order of $g, r, i, z, g-r, r-i, i-z, W1, W2$ and $W1-W2$, for a total of 10 channels. The multi-band imaging of an ELG source is shown in Figure~\ref{fig:image_data}. In the end, our final known dataset comprises 192375 ELG sources, each containing 10-channel images and 85 photometric features. The optical-band images have a resolution of 0.262 arc seconds per pixel, while the infrared-band images have a resolution of 2.75 arc seconds per pixel, with each image having dimensions of 64 $\times$ 64 pixels.

\section{The Method} \label{sec:method}
\begin{figure*}[ht]
    \centering
    \includegraphics[width=0.9\linewidth]{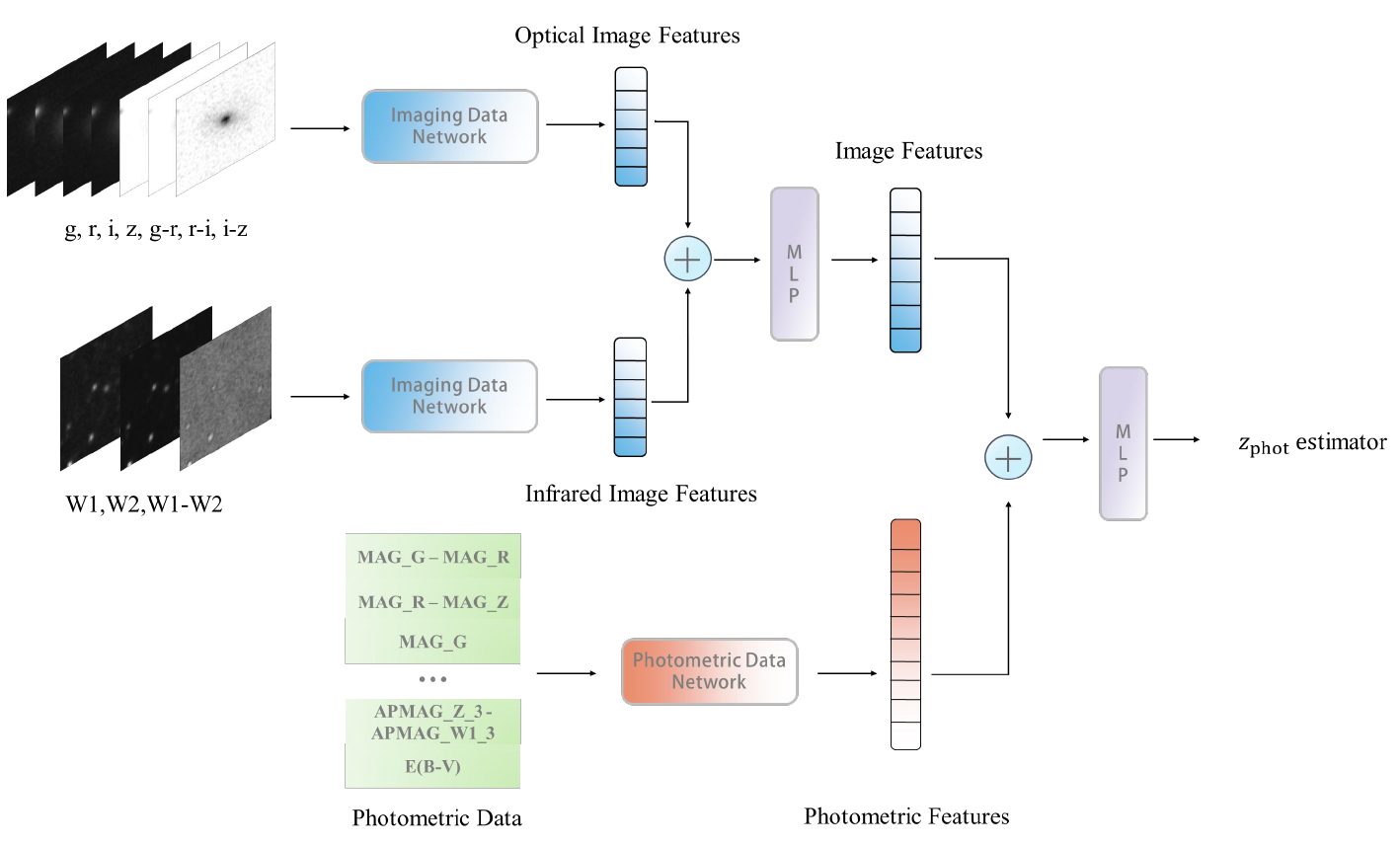}
    \caption{Schematic diagram of the CNN-MLP model. It consists of two distinct networks: the Imaging Data Network, which processes optical- and infrared-band images through two parallel modules, and the Photometric Data Network, which handles the photometric data. The image features and photometric features are concatenated together and passed through a MLP to estimate $z_\mathrm{{phot}}$.}
    \label{fig:model}
\end{figure*}

In this study, we put forward a CNN-MLP model, which combines both image and photometric data to enhance the accuracy of $z_\mathrm{{phot}}$ estimation. The CNN-MLP model architecture, as shown in Figure~\ref{fig:model}, consists of two main components: one is a parallel set of CNN modules, each handling multi-band imaging data from optical and infrared bands. The input data cubes for these modules have shapes of (7, 64, 64) for optical bands and (3, 64, 64) for infrared bands. The other component is an MLP that processes the 85 photometric features. These features from both components are concatenated and subsequently passed through another MLP to estimate $z_\mathrm{{phot}}$ in a classification manner. In the following, we will provide detailed descriptions of the model structure.

\subsection{Imaging data network}
\begin{figure*}
    \centering
    \includegraphics[width=\linewidth]{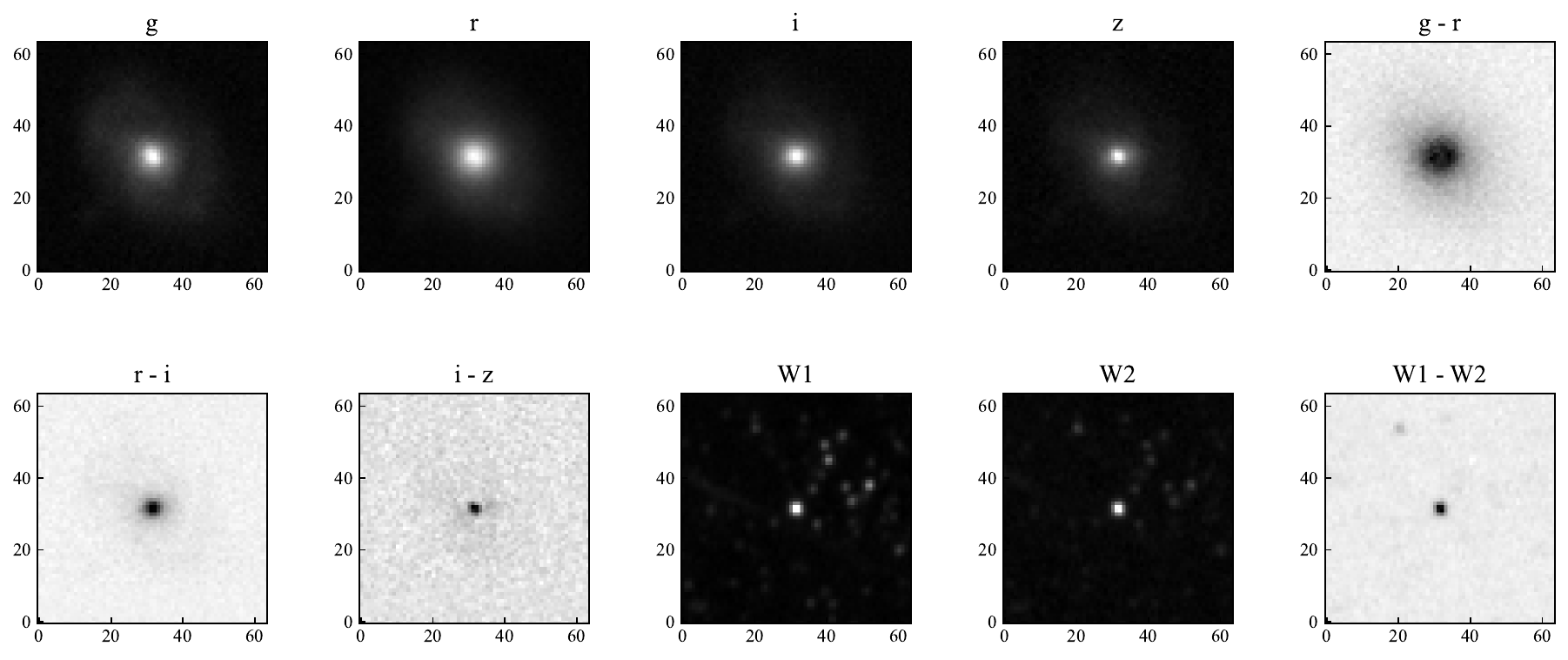}
    \caption{Example multi-band images of a single ELG source with a $z_\mathrm{{spec}}$ of 0.156 from the dataset. The 64 $\times$ 64 imaging data consist of 10 channels, with the bands ordered from optical to infrared. Notable resolution discrepancies are observed between the optical and infrared images.
    }
    \label{fig:image_data}
\end{figure*}

We aim to incorporate infrared-band data from NEOWISE to complement the optical-band information. However, our experiments show that simply concatenating them along the channel axis, forms a datacube of shape (10, 64, 64), leads to performance degradation, with results even worse than using the optical band alone. This can be attributed to the resolution differences between DESI LS and NEOWISE, as illustrated in Figure~\ref{fig:image_data}. Specifically, DESI LS has a resolution of 0.262''/pixel, while NEOWISE has a resolution of 2.75''/pixel. The varying source sizes across different channels may confuse the model regarding morphological features, resulting in ambiguous learned visual features. Therefore, we design two independent CNN modules: one for DESI LS images and another for NEOWISE. Both modules share the same architecture.

The CNN architecture adopted for imaging data closely follows and integrates the designs proposed by \citet{Pasquet19} and \citet{Treyer24}. The network is initialised with a 5$\times$5 convolutional layer followed by an average pooling layer, and is then followed by five inception modules. The inception module is built on the principle of multi-scale feature extraction, wherein convolutional kernels of varying sizes are leveraged to capture distinct spatial hierarchical features within images. Each inception module is structured in two stages. In the first stage, the feature maps are convolved by three 1 $\times$ 1 convolutional layers. The employment of 1 $\times$ 1 convolutions acts as a form of dimensionality reduction, which is essential for managing the computational complexity that arises from the subsequent larger convolutions. In the second stage, feature maps are processed in parallel through a pooling layer and a pair of larger convolutional layers with kernel sizes of 3 $\times$ 3 and 5 $\times$ 5, motivated by the need to capture larger and more complex patterns. By combining features extracted from multiple kernel sizes, the inception module can capture patterns at different resolutions.

Following the inception blocks, we append three additional 3 $\times$ 3 convolutional layers without padding, accompanied by an average pooling layer to further compress the spatial information.
To account for the effect of galactic extinction on the observed imaging data, we incorporate the galactic extinction value E(B-V) at this stage. Specifically, the flattened image features are concatenated with the corresponding E(B-V) value and then passed through a one-layer MLP to better fuse the image features and extinction information, thereby enabling the network to correct for reddening-induced biases in downstream redshift estimation. Through this sequence of modules, a feature representation of the imaging data is obtained.

After the two parallel CNN modules separately process the optical and infrared images, their respective extracted features are flattened and concatenated together. Subsequently, a two-layer MLP is applied to fuse the features from different bands and reduce the dimensionality, thereby producing a feature vector that represents the imaging data.

\subsection{Photometric data network}
Imaging data capture visual features such as shape and structure while photometric data provide precise numerical flux and colour information. They represent different underlying information of the target sources in different modalities, complementing each other in $z_\mathrm{{phot}}$ estimation. As shown in Figure~\ref{fig:model}, we employ a separate MLP to process the photometric data features and extract the information independently. As a result, the overall model architecture can be viewed as a combination of two distinct networks. This structure enables efficient feature extraction from each modality, while avoiding challenges associated with increased model complexity and interpretability.

\subsection{Feature fusion and redshift estimation}\label{sec:feature_fusion}
After being processed by the imaging and photometric data networks, the resulting features are concatenated and passed through an MLP to extract the fused representation for redshift estimation.

The redshift estimation task has been approached using regression, as in \citet{Henghes22}, classification, as in \citet{Pasquet19}, or a combination of both, as in \citet{Treyer24}. 
In a regression-based approach, the model outputs a single value representing the estimated redshift, and training is performed by minimising the difference between the predicted and true redshift values, quantified by mean square error (MSE) or mean absolute error 
(MAE). 
 But in a classification method, the known redshift range is divided into a series of narrow, mutually exclusive bins, and a classifier is trained to assign each sample to one of these bins. The classifier outputs confidence scores indicating the likelihood of the sample belonging to each bin, which can be transformed into probability values using the softmax function. This provides a reliable estimate of the $z_\mathrm{{phot}}$ probability density function (PDF). Finally, based on the obtained PDF, the photometric redshift can be calculated using Equation ~\ref{eq:point_estimate}.
 
\begin{equation}
    z_\mathrm{{phot}} = \sum_{k=1}^{N_c} z_k P(z_k)
    \label{eq:point_estimate}
\end{equation}
where $N_c$ is the number of bins, $z_k$ is the midpoint value of the $k$-th redshift bin, and $P(z_k)$ denotes the probability that the sample's  redshift falls into the $k$-th bin.
Under the classification schema, the redshift label is one-hot encoded to represent the corresponding bin, and
the classifier is trained using the cross-entropy loss function.

We conduct an experiment to compare the performance of the two approaches, as detailed in \ref{sec:class_vs_regre}. The results show that the classification method performs better than the regression method. Therefore, we adopted the classification method to construct the redshift estimator.

\section{Experiment}
\subsection{Metrics} \label{metrics}

To evaluate the performance of $z_\mathrm{{phot}}$ estimation, we consider the following commonly used metrics:
\begin{itemize}
    \item The normalised residuals $\Delta z$, following \citet{residuals2000}:
    \begin{equation}
    \Delta z = \frac{z_\mathrm{{spec}} - z_\mathrm{{phot}}}{1+z_\mathrm{{spec}}}
\end{equation}
\item The prediction Bias: the median of the normalised residuals. We adopt the median-based definition following \citet{Roster24}, as the median offers a robust measure of central tendency that is less sensitive to extreme values;
\item The normalised median absolute deviation ($\sigma_{\mathrm{NMAD}}$),  following \citet{EAZY}, as it is less sensitive to outliers:
\begin{equation}
    \sigma_{\mathrm{NMAD}} = 1.48 \times \mathrm{median}|\Delta z - \mathrm{median}(\Delta z)|
\end{equation}
\item The outlier fraction $\eta$, following \citet{outlier2012}:

\begin{equation}
    \eta = \frac{N_{|\Delta z|>0.15}}{N_{\rm total}}
\end{equation}

where $N_{|\Delta z|>0.15}$ denotes the number of samples for which
$|\Delta z|$ is larger than 0.15. $N_{\rm total}$ is the total number of samples.

\end{itemize}

\subsection{Experiment setup and hyperparameter tuning}
The CNN-MLP approach is implemented using the Pytorch framework \citep{pytorch} and all experiments are performed in a high performance computing cluster provided by the National Astronomical Data Center (NADC; \citealt{lich2017}). 

The known ELG sample is partitioned into training, validation, and test sets in an approximate ratio of 75:5:20, corresponding to 144663, 9237 and 38475 samples, respectively. Moreover, the redshift distribution of training, validation, and test sets is consistent. In addition, to reduce the effect of galaxy orientation on our imaging data model, we augment the training set by applying random flips and rotations (in 90$^{\circ}$ increments) to the images.

A number of hyperparameters play a critical role in determining the network architecture and influencing convergence of the training. To  optimise model performance, we utilise Weights \& Bias sweep tool \citep{wandb} to perform a hyperparameter search based on the validation set, aiming to identify the optimal combination that yields the best results. The key parameters with the most significant impact are outlined below:
\begin{itemize}
    \item \textbf{Imaging data network architecture:}
    \begin{itemize}
        \item Number of convolutional layers after Inception modules: [1, 2, 3]
        \item Number of channels in each convolutional layer after Inception modules: [64, 96, 128]
    \end{itemize}
    
    \item \textbf{Photometric data network architecture:}
    \begin{itemize}
        \item Number of MLP layers: [1, 2, 3]
        \item Number of neurons in each MLP layer: [512, 1024, 2048]
    \end{itemize}
    
    \item \textbf{Training configuration:}
    \begin{itemize}
        \item Dropout rate: [0.1, 0.3, 0.5]
        \item Activation function: [ReLU, GELU, LeakyReLU]
        \item Initial learning rate: [1 $\times$ $10^{-5}$ -- 1$\times 10^{-3}$]
        \item Step size in StepLR learning rate scheduler: [5, 10, 15]
        \item Gamma in StepLR scheduler (learning rate decay factor): [0.1, 0.3, 0.5]
        \item Weight decay in optimizer: [1$\times10^{-6}$ -- 1$\times 10^{-2}$]
        \item Batch size: [32--128]
        \item Number of training epochs: [30--50]
    \end{itemize}
\end{itemize}

\begin{table*}[ht]
\centering
\caption{Optimal hyperparameters for CNN-MLP model and its sub-networks. The imaging data network (CNN) and the photometric data network (MLP) are first trained separately, followed by joint fine-tuning of the full CNN-MLP model.}
\begin{tabular}{cccc}
\hline
\textbf{Hyperparameter} & \textbf{Imaging data network} & \textbf{Photometric data network} & \textbf{CNN-MLP} \\
\hline
Number of convolutional layers after Inception modules & 3 & - & - \\
Number of channels in each convolutional layer after Inception modules & 96 & - & - \\
Number of MLP layers & - & 2 & - \\
Number of neurons in each MLP layer & - & 512 & - \\
Number of bins in classification-based estimation & - & - & 770 \\
Dropout rate & 0.3 & 0.1 & - \\
Activation function & ReLU & GELU & ReLU \\
Initial learning rate & 0.0005 & 0.0009 & 0.0001 \\
Step size in StepLR scheduler & 15 & 10 & 10 \\
Gamma in StepLR scheduler & 0.1 & 0.1 & 0.1 \\
Weight decay in optimizer & 1e-3 & 1e-5 & 1e-5 \\
Number of training epochs & 50 & 30 & 30 \\
Batch size & 32 & 32 & 32 \\
\hline
\end{tabular}
\label{tab:hyperparameter}
\end{table*}

In addition to the hyperparameter values obtained from the search, there are several other model hyperparameter settings as follows: Xavier initialization for the network's weights is employed, and the bias is set to a constant value of 0.1. The Adam optimizer is used during training. The number of bins in the final estimation, as described in Section \ref{sec:feature_fusion}, is 770, covering the redshift range from 0 to 3.85, with each bin having a width of $\Delta z = 0.005$.

Rather than training the entire CNN-MLP model in an end-to-end manner from the beginning, we adopt a staged training strategy. Specifically, the imaging data network and the photometric data network are first trained and fine-tuned separately to obtain optimal parameters for each modality. These pretrained weights are then loaded into the full CNN-MLP, followed by end-to-end fine-tuning with a small learning rate.
This staged approach improves both efficiency and performance. By first fine-tuning each modality-specific module independently, the feature extraction capacity of each network is maximised, resulting in more stable training and faster convergence. Building upon pretrained weights, the fusion and redshift estimation layers can then be optimised at minimal cost. The final joint fine-tuning step allows the two modalities to adapt cooperatively, enhancing feature complementarity and mitigating potential distributional biases introduced during separate training.

The optimal hyperparameters for each network and the joint fine-tuning stage are shown in Table~\ref{tab:hyperparameter}.
Subsequently, the model is built using these hyperparameters to construct the final CNN-MLP model.

\section{Results and Discussion} \label{sec:performance}
\subsection{Results}
To highlight the superiority of the CNN-MLP architecture, we construct two single-modality baseline models by removing either the CNN or MLP component: one retains only the MLP for processing photometric data, and the other retains only the CNN for imaging data. Their performances are then compared with the full model on the same dataset.

Table \ref{tab:main_res} presents the results of the performance comparison. The use of multimodal data by CNN-MLP yields the best performance across all three metrics, with significant improvements of 12.5\% and 14.6\% in $\sigma_{\mathrm{NMAD}}$ compared to the MLP using only photometric data and CNN using only imaging data, respectively. 

\begin{table}[ht]
    \centering
    \caption{The performance comparison of different models.}
    \begin{tabular}{ccccc}
    \hline
         Method& Data & Bias & $\sigma_{\mathrm{NMAD}}$  & $\eta$ \\
         \hline
        CNN-MLP & Multimodal data & 0.0002 & \textbf{0.0140} & \textbf{0.0257}\\
         MLP & Photometric data & \textbf{0.0000} & 0.0160 & 0.0284 \\
         CNN & Imaging data & 0.0001 & 0.0164 & 0.0316\\
         \hline
    \end{tabular}
    \label{tab:main_res}
\end{table}

These results can be attributed to several factors. First, imaging data not only provide flux and colour, which are directly related to redshift estimation, but also contain additional information such as the galaxy's apparent size, extended features, morphology, and spatial structure. These characteristics, which cannot be captured in photometric data alone, are also valuable for predicting redshifts. However, there are limitations when solely using imaging data. Galaxies vary in size, and the limited field of view in images may lead to contamination from background sources, introducing additional flux that can affect the redshift estimation of the central galaxy. 

On the other hand, while photometric data are typically derived from images, they undergo careful photometric processing and extinction corrections, making the information on flux and colour more accurate. The flux measurements include both model flux and aperture flux, providing richer information on luminosity distribution. This is why traditional machine learning methods can achieve relatively good performance with photometric data alone.

\begin{figure*}	[ht]
		\begin{minipage}[t]{0.32\textwidth}
			\centering
                \includegraphics[width=\linewidth]{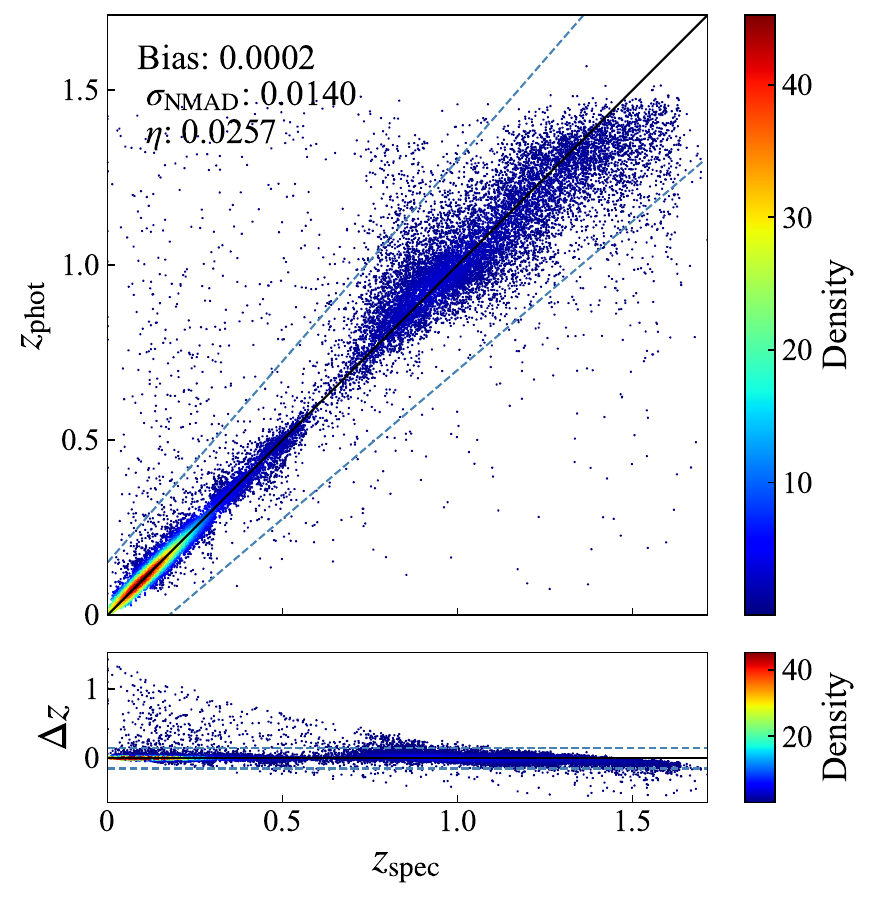}
		\end{minipage}%
	\begin{minipage}[t]{0.32\textwidth}
		\centering
		  \includegraphics[width=\linewidth]{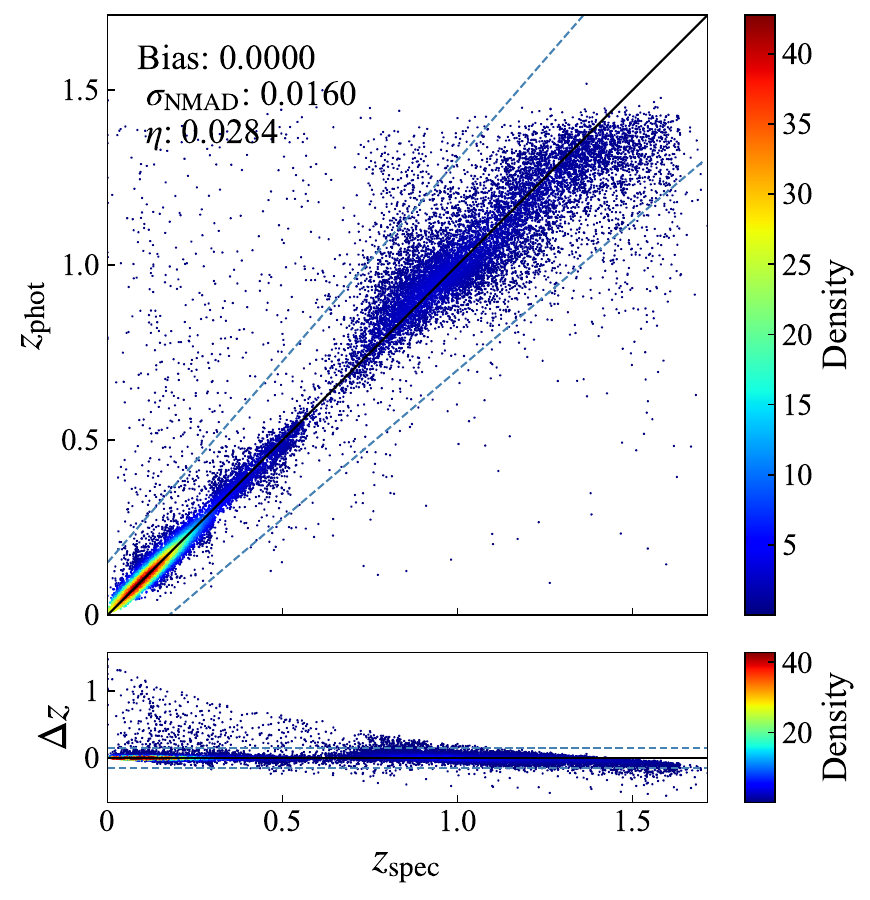}
	\end{minipage}%
		\begin{minipage}[t]{0.32\textwidth}
			\centering
                 \includegraphics[width=\linewidth]{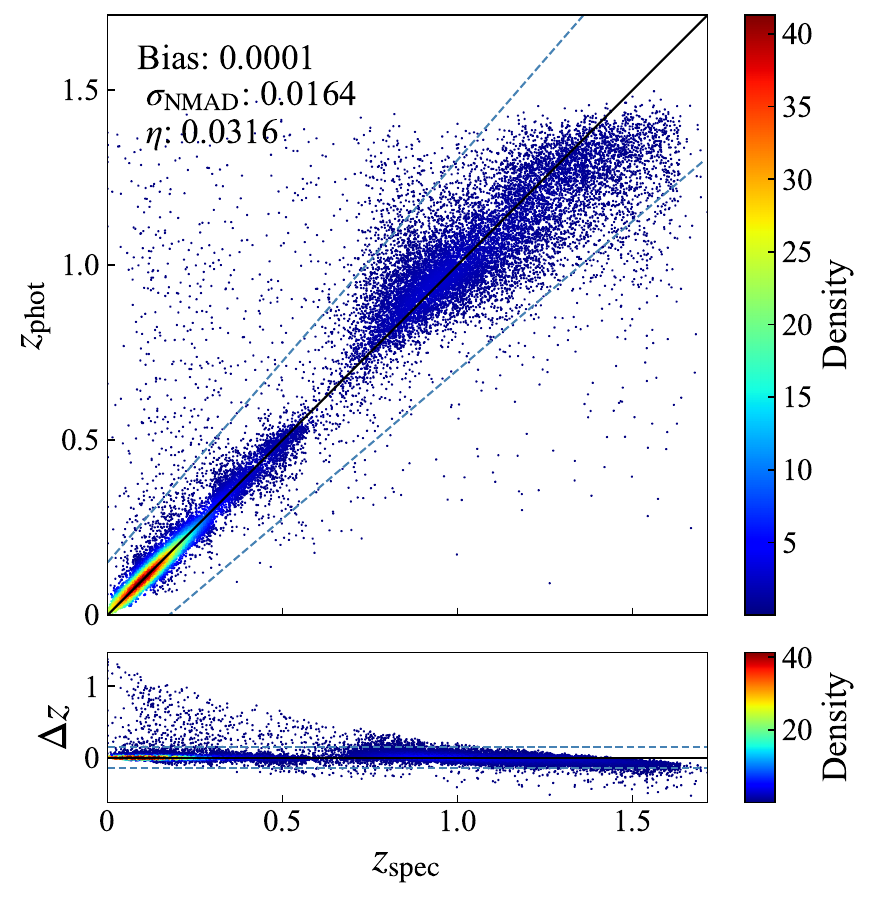}
		\end{minipage}%
    
    \caption{$z_\mathrm{{phot}}$ computed with CNN-MLP, MLP and CNN. Top:$z_\mathrm{{phot}}$ versus $z_\mathrm{{spec}}$. Bottom: normalised residuals across the redshift range. The black solid line represents the one-to-one relation with no residuals, while the blue dashed lines correspond to $z_\mathrm{{phot}}$ at $\pm$ 0.15(1+$z_\mathrm{{spec}}$). Sources outside the dashed lines are identified as outliers. The colour intensity indicates the density of samples. Left panel: CNN-MLP model, middle panel: MLP model, right panel: CNN model.}
    \label{fig:three_models_res}
\end{figure*}

The distribution of $z_\mathrm{{phot}}$ versus $z_\mathrm{{spec}}$ for CNN-MLP, MLP, and CNN is shown in Figure~\ref{fig:three_models_res}. The densities for all three models are largely concentrated along the $z_\mathrm{{phot}} = z_\mathrm{{spec}}$ lines, indicating high prediction accuracy. However, the CNN model exhibits an increased dispersion around $z_\mathrm{{spec}}$, consistent with the higher $\sigma_{\mathrm{NMAD}}$ reported in Table~\ref{tab:main_res} and the trends observed across different redshift ranges discussed in Section~\ref{sec:redshift_range}.

In addition to the excellent performance of the CNN-MLP model, the CNN model using only imaging data achieves solid performance. As a result, it is particularly promising for parameter estimation directly using imaging data from future multi-band imaging surveys, such as CSST and LSST.

\subsection{Performance in different redshift and magnitude ranges} \label{sec:redshift_range}
\begin{figure}[h]
    \centering
    \includegraphics[width=0.9\linewidth]{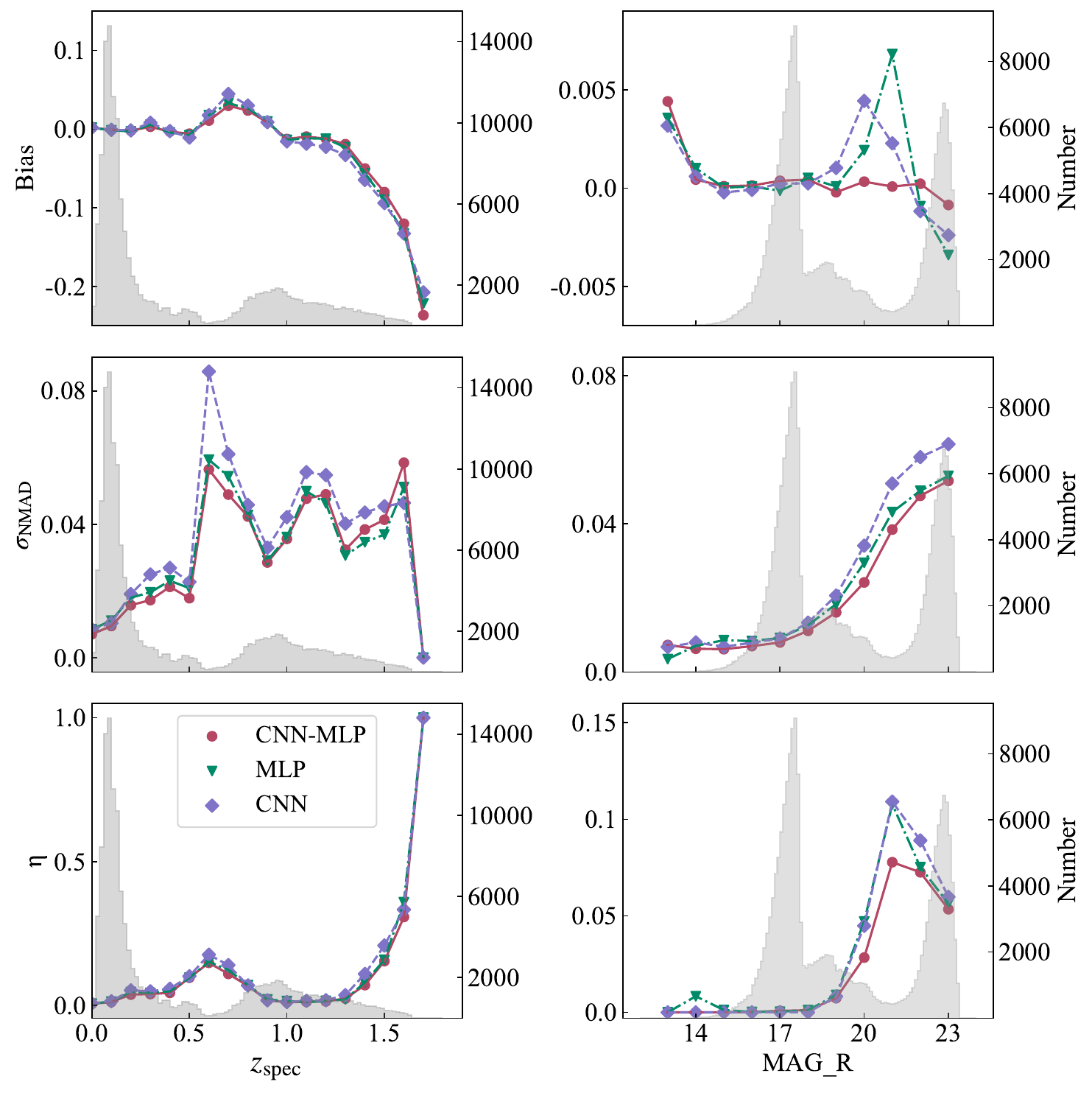}
    \caption{Performance of different models as a function of $z_\mathrm{spec}$ and $r$-band model magnitude. Gray histograms in the background show the distributions of $z_\mathrm{spec}$ and $r$-band model magnitudes for the sample.}
    \label{fig:metrics_z_mag}
\end{figure}

The performance of the three models (CNN-MLP, MLP and CNN) as a function of redshift and $r$-band magnitude is shown in Figure~\ref{fig:metrics_z_mag}. 
The metrics analysed include bias, $\sigma_{\mathrm{NMAD}}$ and $\eta$. 

It can be observed from Figure~\ref{fig:metrics_z_mag} that all three metrics exhibit a consistent trend.
For redshifts below 0.5, where sample density is highest, redshift predictions exhibit the greatest accuracy. In contrast, for redshifts above 1.5, where sample density is lower, prediction accuracy diminishes significantly. The moderate degradation in metrics within the 0.5–1.0 range can be attributed to the relatively sparse sample density in this interval. The observed performance degradation due to uneven sample distribution is a well-known challenge in machine learning: models tend to perform better on classes with abundant samples while generalising less effectively to those with fewer samples. Future work should focus on enhancing performance in the sparsely sampled high-redshift regime by employing techniques such as data augmentation, synthetic data generation, or other strategies to address the issue of data imbalance.

As for magnitude, it is clear that redshift estimation accuracy is generally higher in the bright magnitude region compared to the faint magnitude region, as observed across all three metrics. The CNN-MLP model consistently achieves the best performance across nearly all magnitude intervals. This suggests that the CNN-MLP model effectively integrates limited information to optimise redshift estimation accuracy for faint sources, even when the morphological and photometric information is not highly precise.

In addition, we apply a two-part model to assess whether training separate models for bright and faint sources could yield an optimal solution. Specifically, the two-part model is trained independently for both bright ($r$-band model magnitude < 21.5) and faint ($r$-band model magnitude > 21.5) data groups, while the previous model, trained on the entire dataset, is referred to as the one-part model.
The performance of bright and faint sources with the one-part model and the two-part model is summarised in Table~\ref{tab:bright_faint_res}. 

For the bright sources, the two-part model achieves nearly identical performance to the one-part model. However, for the faint sources, it obtains a slight performance deterioration compared to the one-part model. This phenomenon can be attributed to two factors: firstly, the data from bright sources are more easily leveraged to extract effective information, whereas faint sources are both fewer in number and intrinsically dim, making it inherently challenging to extract accurate and reliable information and thus limiting the model’s ability to learn sufficiently from them. This issue is mitigated when training on the entire dataset. Secondly, there are some shared features and patterns between bright and faint sources. These features that may not be readily extracted from faint images could still be learned from the intrinsic similarities shared with bright and high-quality images. Training on the entire dataset allows the model to fully leverage those shared features, whereas training exclusively by faint sources restricts this ability. These findings suggest that when using CNN-MLP, incorporating information from both bright and faint sources with the one-part model may provide a more balanced and robust solution for redshift prediction.

\begin{table}[]
\centering
\caption{Performance of bright and faint sources using the two-part model and one-part model.}
\begin{tabular}{ccccc}
\toprule
Model & Sources & Bias & $\sigma_{\mathrm{NMAD}}$ & $\eta$ \\
\hline
 & MAG\_R < 21.5 & 0.0000 & 0.0089 & 0.0044 \\
Two-part Model & MAG\_R > 21.5 & 0.0010 & 0.0494 & 0.0715 \\
& All & 0.0001 & 0.0139 & 0.0270 \\
\hline
 & MAG\_R < 21.5 & 0.0003 & 0.0090 & 0.0041 \\
One-part Model & MAG\_R > 21.5 & 0.0000 & 0.0479 & 0.0691 \\
& All & 0.0002 & 0.0140 & 0.0257 \\
\hline
\end{tabular}

\label{tab:bright_faint_res}
\end{table}

\subsection{Performance for different ELG types}
\begin{figure}[ht]
    \centering
    \includegraphics[width=0.9\textwidth]{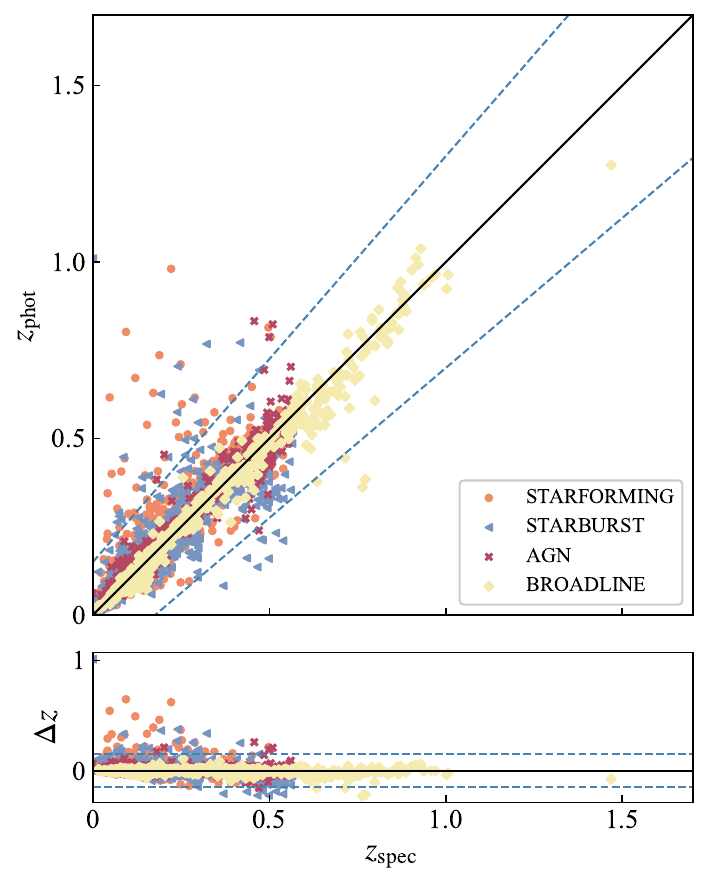}
    \caption{Performance for different galaxy types. Top:$z_\mathrm{{phot}}$ versus $z_\mathrm{{spec}}$. Bottom: normalised residuals across the redshift range.}
    \label{fig:galaxy_types}
\end{figure}

\begin{figure*}[ht]
    \centering
    \includegraphics[width=0.9\linewidth]{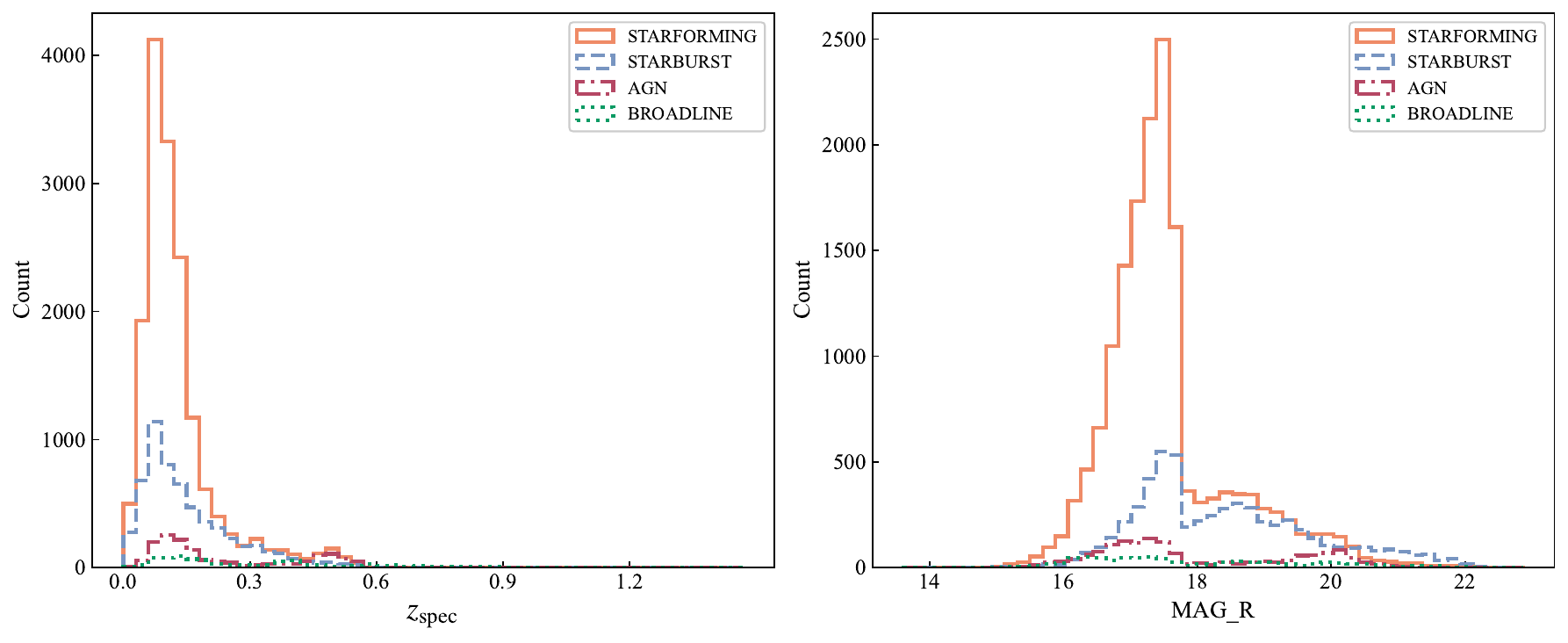}
    \caption{Distributions of $z_\mathrm{{spec}}$ and $r$-band model magnitude for different galaxy types in the test set.}
    \label{fig:galaxy_hist}
\end{figure*}

The ELG types in the SDSS include starforming, starburst, AGN, and broadline galaxies, as described in Section \ref{sec:spec_redshift}. To gain a deeper understanding of performance across different ELG types, we select different types of SDSS ELGs from the test set and calculate the three metrics for each ELG type group. The results are presented in Table \ref{tab:galaxy_types} and Figure~\ref{fig:galaxy_types}. As shown in Figure~\ref{fig:galaxy_types}, the majority of SDSS galaxy sources are concentrated in the low-redshift region, in contrast to the broader distribution in the full test set. Consequently, the redshift estimation performance on the SDSS dataset is superior to that on the entire dataset, as discussed in Section \ref{sec:redshift_range}. Additionally, Table \ref{tab:galaxy_types} reveals minor variations in performance across different ELG types. These discrepancies can largely be attributed to differences in their respective redshift and brightness distributions. Figure~\ref{fig:galaxy_hist} illustrates these distributions, showing that the best-performing starforming galaxies are predominantly located at lower redshifts and are generally brighter. In contrast, AGN and broadline galaxies, which perform relatively poorly, exhibit broader redshift distributions and more balanced magnitude distributions, with comparable numbers of bright and faint sources.
As discussed in Sections \ref{sec:redshift_range} , $z_\mathrm{{phot}}$ prediction becomes more challenging for galaxies at higher redshifts and with lower brightness. Moreover, the sample size of each galaxy type also plays a role in model performance: starforming galaxies constitute more than half of the training dataset, which facilitates more effective model learning for this population.

\begin{table}[ht]
\centering
\caption{Performance for different galaxy types.}
\begin{tabular}{lcccc}
\toprule
Galaxy Type & No. of sources & Bias & $\sigma_{\mathrm{NMAD}}$ & $\eta$ \\
\hline
Starforming & 15944 & 0.0003 & 0.0083 & 0.0023 \\
Starburst & 5738 & 0.0003 & 0.0101 & 0.0052 \\
AGN & 1512 & 0.0030 & 0.0134 & 0.0040 \\
Broadline & 803 & -0.0027 & 0.0103 & 0.0050 \\
All & 23997 & 0.0003 & 0.0090 & 0.0032 \\
\hline
\end{tabular}

\label{tab:galaxy_types}
\end{table}

\subsection{Outliers} \label{outliers}
\begin{figure}[ht]
    \centering
    \includegraphics[width=0.9\linewidth]{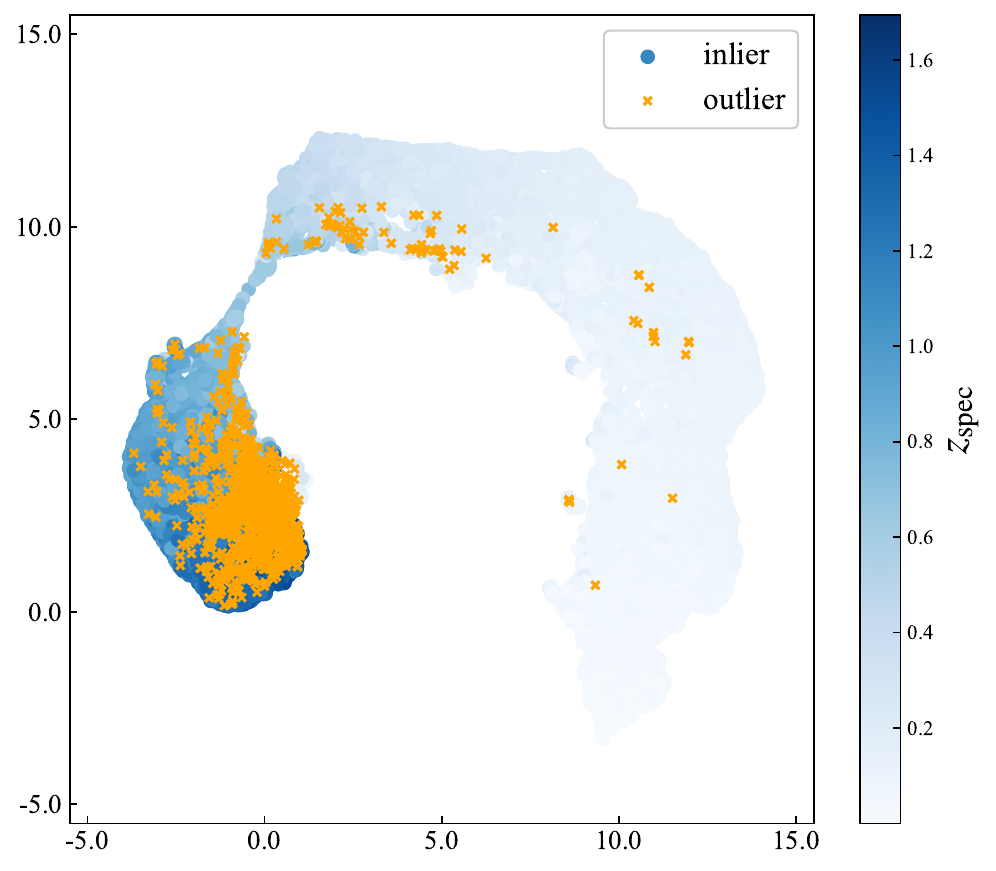}
    \caption{Two-dimensional UMAP projection of the 1024-dimensional network output, colour-coded by $z_\mathrm{{spec}}$. Outliers are marked as orange crosses.}
    \label{fig:outlier_umap}
\end{figure}

As discussed in Section \ref{metrics}, samples with $|\Delta z|$ larger than 0.15 are classified as outliers, while those with smaller values are considered inliers. To investigate the differences between inliers and outliers predicted by the model, Uniform Manifold Approximation and Projection \citep[UMAP;][]{UMAP} is applied to reduce the high-dimensional feature space to two dimensions for visualization, as shown in Figure~\ref{fig:outlier_umap}.

Figure~\ref{fig:outlier_umap} shows the distribution of inliers, colour-coded by their matched $z_\mathrm{{spec}}$ values, alone with the outliers (represented as orange crosses) in the projected space. This analysis aims to determine whether the two groups exhibit distinguishable clustering or separation patterns, potentially revealing underlying feature characteristics that contribute to prediction errors. For the majority of samples, the model has effectively learned redshift representations, with well-defined clusters corresponding to both low- and high-redshift sources. However, the presence of outlier clusters suggests that certain intrinsic features of these sources lead to prediction errors. Notably, many outliers are concentrated in the high-redshift region, indicating that their learned representations are similar to those of high-redshift galaxies, even though they belong to the low-redshift group. This overlap highlights potential degeneracies in the dataset, where sources from both high and low-redshift populations share similar characteristics, complicating the model’s ability to distinguish between them.

\begin{figure}
    \centering
    \includegraphics[width=0.9\linewidth]{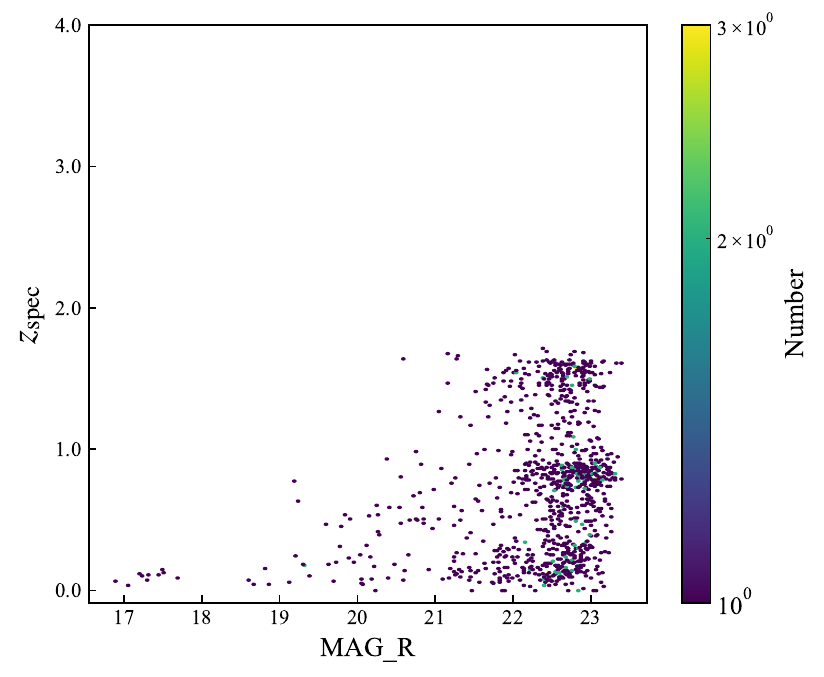}
    \caption{Distribution of outliers in the $z_\mathrm{spec}$ vs. $r$-band magnitude diagram.}
    \label{fig:outlier_z_r_dist}
\end{figure}

The distribution of outliers in the $z_\mathrm{spec}$ versus $r$-band magnitude space is described in Figure~\ref{fig:outlier_z_r_dist}.  In addition to the challenges associated with faint high-redshift sources discussed in \ref{sec:redshift_range}, we also observe a notable concentration of outliers in the lower-right region, corresponding to faint sources at low redshifts. This pattern likely arises due to the under-representation of such galaxies in the training set, as illustrated in Figure~\ref{fig:redshift_r}. This imbalance introduces a training bias, where the 
model tends to associate faintness with higher redshifts. As a result, faint low-redshift galaxies are more prone to being misclassified, contributing to the observed outliers.

We also conclude that accurately estimating $z_\mathrm{phot}$ for faint galaxies—whether at high or low redshifts—is particularly challenging due to several compounding factors. First, a low signal-to-noise ratio (SNR) hampers the precision of redshift measurements. We examined the SNR distributions of inliers and outliers and found that outliers exhibit significantly lower SNRs across all photometric bands. Furthermore, the limited depth of the survey restricts both the detectability and measurement quality of faint sources, which in turn contributes to low SNR and reduced estimation accuracy. In addition, cosmic dust and extinction further degrade the already weak photometric images, making them even less reliable. Finally, contamination from nearby objects introduces extra fluxes, affecting photometric measurements and reducing the accuracy of $z_\mathrm{{phot}}$ estimation.

\section{Conclusion} \label{sec:conclusions}
In this paper, we propose a multimodal deep learning model CNN-MLP that integrates imaging and photometric data to enhance the redshift estimation of ELGs. Compared to single-modal approaches, the multimodal model achieves better performance, with the bias $=0.0002$, $\sigma_{\mathrm{NMAD}}$ = 0.0140 and outlier fraction $\eta$ = 0.0257. This result demonstrates that visual and photometric information complement each other, significantly improving $z_\mathrm{{phot}}$ estimation.
Additionally, we find that model performance degrades with increasing redshift and fainter magnitudes, highlighting the inherent difficulties in extracting information and accurately estimating photometric redshifts for high-redshift and faint sources. This degradation also reflects the impact of data imbalance, as such sources are underrepresented in the dataset. Moreover, we compare photometric redshift estimation performance across different ELG types, showing that starforming galaxies—typically brighter, at lower redshift, and more numerous—exhibit better performance than other ELG types. 

With the help of the CNN-MLP model, we can estimate the redshifts of DESI's ELG targets to further optimise the selection of observation targets. Additionally, this model may be applied to ongoing and upcoming surveys such as Euclid, LSST, and CSST.

\paragraph{Acknowledgments}
We appreciate the insightful comments provided by the referee, which have been instrumental in improving our paper. This work is supported by National Natural Science Foundation of China
(NSFC; grant Nos. 12373110, 12273076, 12403102, 12103070, 12273077, 12133001), the National
Key Research and Development Program of China (2022YFF0711500), Strategic Priority Research Program of the Chinese Academy of Sciences (XDB0550101). Data resources are supported by China National Astronomical Data Center (NADC), CAS Astronomical Data Center and Chinese Virtual Observatory (China-VO). This work is supported by Astronomical Big Data Joint Research Center, co-founded by National Astronomical Observatories, Chinese Academy of Sciences and Alibaba Cloud.\\

The Legacy Surveys consist of three individual and complementary projects: the Dark Energy Camera Legacy Survey (DECaLS; Proposal ID \#2014B-0404; PIs: David Schlegel and Arjun Dey), the Beijing-Arizona Sky Survey (BASS; NOAO Prop. ID \#2015A-0801; PIs: Zhou Xu and Xiaohui Fan), and the Mayall z-band Legacy Survey (MzLS; Prop. ID \#2016A-0453; PI: Arjun Dey). DECaLS, BASS and MzLS together include data obtained, respectively, at the Blanco telescope, Cerro Tololo Inter-American Observatory, NSF’s NOIRLab; the Bok telescope, Steward Observatory, University of Arizona; and the Mayall telescope, Kitt Peak National Observatory, NOIRLab. Pipeline processing and analyses of the data were supported by NOIRLab and the Lawrence Berkeley National Laboratory (LBNL). The Legacy Surveys project is honored to be permitted to conduct astronomical research on Iolkam Du’ag (Kitt Peak), a mountain with particular significance to the Tohono O’odham Nation.\\

NOIRLab is operated by the Association of Universities for Research in Astronomy (AURA) under a cooperative agreement with the National Science Foundation. LBNL is managed by the Regents of the University of California under contract to the U.S. Department of Energy.\\

This project used data obtained with the Dark Energy Camera (DECam), which was constructed by the Dark Energy Survey (DES) collaboration. Funding for the DES Projects has been provided by the U.S. Department of Energy, the U.S. National Science Foundation, the Ministry of Science and Education of Spain, the Science and Technology Facilities Council of the United Kingdom, the Higher Education Funding Council for England, the National Center for Supercomputing Applications at the University of Illinois at Urbana-Champaign, the Kavli Institute of Cosmological Physics at the University of Chicago, Center for Cosmology and Astro-Particle Physics at the Ohio State University, the Mitchell Institute for Fundamental Physics and Astronomy at Texas A\&M University, Financiadora de Estudos e Projetos, Fundacao Carlos Chagas Filho de Amparo, Financiadora de Estudos e Projetos, Fundacao Carlos Chagas Filho de Amparo a Pesquisa do Estado do Rio de Janeiro, Conselho Nacional de Desenvolvimento Cientifico e Tecnologico and the Ministerio da Ciencia, Tecnologia e Inovacao, the Deutsche Forschungsgemeinschaft and the Collaborating Institutions in the Dark Energy Survey. The Collaborating Institutions are Argonne National Laboratory, the University of California at Santa Cruz, the University of Cambridge, Centro de Investigaciones Energeticas, Medioambientales y Tecnologicas-Madrid, the University of Chicago, University College London, the DES-Brazil Consortium, the University of Edinburgh, the Eidgenossische Technische Hochschule (ETH) Zurich, Fermi National Accelerator Laboratory, the University of Illinois at Urbana-Champaign, the Institut de Ciencies de l’Espai (IEEC/CSIC), the Institut de Fisica d’Altes Energies, Lawrence Berkeley National Laboratory, the Ludwig Maximilians Universitat Munchen and the associated Excellence Cluster Universe, the University of Michigan, NSF’s NOIRLab, the University of Nottingham, the Ohio State University, the University of Pennsylvania, the University of Portsmouth, SLAC National Accelerator Laboratory, Stanford University, the University of Sussex, and Texas A\&M University.\\

BASS is a key project of the Telescope Access Program (TAP), which has been funded by the National Astronomical Observatories of China, the Chinese Academy of Sciences (the Strategic Priority Research Program “The Emergence of Cosmological Structures” Grant \# XDB09000000), and the Special Fund for Astronomy from the Ministry of Finance. The BASS is also supported by the External Cooperation Program of Chinese Academy of Sciences (Grant \# 114A11KYSB20160057), and Chinese National Natural Science Foundation (Grant \# 12120101003, \# 11433005).

The Legacy Survey team makes use of data products from the Near-Earth Object Wide-field Infrared Survey Explorer (NEOWISE), which is a project of the Jet Propulsion Laboratory/California Institute of Technology. NEOWISE is funded by the National Aeronautics and Space Administration.

The Legacy Surveys imaging of the DESI footprint is supported by the Director, Office of Science, Office of High Energy Physics of the U.S. Department of Energy under Contract No. DE-AC02-05CH1123, by the National Energy Research Scientific Computing Center, a DOE Office of Science User Facility under the same contract; and by the U.S. National Science Foundation, Division of Astronomical Sciences under Contract No. AST-0950945 to NOAO.\\

DESI construction and operations is managed by the Lawrence Berkeley National Laboratory. This research is supported by the U.S. Department of Energy, Office of Science, Office of High-Energy Physics, under Contract No. DE–AC02–05CH11231, and by the National Energy Research Scientific Computing Center, a DOE Office of Science User Facility under the same contract. Additional support for DESI is provided by the U.S. National Science Foundation, Division of Astronomical Sciences under Contract No. AST-0950945 to the NSF’s National Optical-Infrared Astronomy Research Laboratory; the Science and Technology Facilities Council of the United Kingdom; the Gordon and Betty Moore Foundation; the Heising-Simons Foundation; the French Alternative Energies and Atomic Energy Commission (CEA); the National Council of Science and Technology of Mexico (CONACYT); the Ministry of Science and Innovation of Spain, and by the DESI Member Institutions. The DESI collaboration is honored to be permitted to conduct astronomical research on Iolkam Du’ag (Kitt Peak), a mountain with particular significance to the Tohono O’odham Nation.\\

Funding for the Sloan Digital Sky 
Survey IV has been provided by the 
Alfred P. Sloan Foundation, the U.S. 
Department of Energy Office of 
Science, and the Participating 
Institutions.

SDSS-IV acknowledges support and 
resources from the Center for High 
Performance Computing  at the 
University of Utah. The SDSS 
website is www.sdss4.org.

SDSS-IV is managed by the 
Astrophysical Research Consortium 
for the Participating Institutions 
of the SDSS Collaboration including 
the Brazilian Participation Group, 
the Carnegie Institution for Science, 
Carnegie Mellon University, Center for 
Astrophysics | Harvard \& 
Smithsonian, the Chilean Participation 
Group, the French Participation Group, 
Instituto de Astrof\'isica de 
Canarias, The Johns Hopkins 
University, Kavli Institute for the 
Physics and Mathematics of the 
Universe (IPMU) / University of 
Tokyo, the Korean Participation Group, 
Lawrence Berkeley National Laboratory, 
Leibniz Institut f\"ur Astrophysik 
Potsdam (AIP),  Max-Planck-Institut 
f\"ur Astronomie (MPIA Heidelberg), 
Max-Planck-Institut f\"ur 
Astrophysik (MPA Garching), 
Max-Planck-Institut f\"ur 
Extraterrestrische Physik (MPE), 
National Astronomical Observatories of 
China, New Mexico State University, 
New York University, University of 
Notre Dame, Observat\'ario 
Nacional / MCTI, The Ohio State 
University, Pennsylvania State 
University, Shanghai 
Astronomical Observatory, United 
Kingdom Participation Group, 
Universidad Nacional Aut\'onoma 
de M\'exico, University of Arizona, 
University of Colorado Boulder, 
University of Oxford, University of 
Portsmouth, University of Utah, 
University of Virginia, University 
of Washington, University of 
Wisconsin, Vanderbilt University, 
and Yale University.\\

Guoshoujing Telescope (the Large Sky Area Multi-Object Fiber Spectroscopic Telescope LAMOST) is a National Major Scientific Project built by the Chinese Academy of Sciences. Funding for the project has been provided by the National Development and Reform Commission. LAMOST is operated and managed by the National Astronomical Observatories, Chinese Academy of Sciences.\\

This research has made use of the SIMBAD database,
operated at CDS, Strasbourg, France.

\paragraph{Data Availability Statement}
The data from DESI LS10 are available at \url{https://portal.nersc.gov/cfs/cosmo/data/legacysurvey/dr10/}.

\bibliography{ref}


\appendix
\section{Photometric features}
\label{sec:feature_selection}
Photometric data provide rich information on galaxy luminosities across multiple bands. To fully exploit this information for $z_\mathrm{{phot}}$ estimation, we construct a set of 97 features by combining the model magnitudes with the aperture magnitudes listed in Table \ref{tab:catalog_param}, and computing various differences between them. Specifically, the features include:
\begin{itemize}[leftmargin=0.5pt]
    \item Model magnitudes in all bands;
    \item Colours derived from model magnitudes, computed both between adjacent and non-adjacent bands (e.g., $MAG\_G - MAG\_R$, $MAG\_G - MAG\_I$, $MAG\_G - MAG\_Z$)
    \item Differences between model magnitudes and the smallest aperture magnitude in the corresponding bands (e.g., $MAG\_G - APMAG\_G\_1$, $MAG\_R - APMAG\_R\_1$);
    \item Colours
     between adjacent bands measured within the same aperture size (e.g., $APMAG\_G\_1 - APMAG\_R\_1$, $APMAG\_R\_1 - APMAG\_I\_1$).
\end{itemize}

\subsection{Selection of photometric features}
To identify the most informative features for $z_\mathrm{{phot}}$ estimation, we employ the Permutation Feature Importance (PFI) method. PFI quantifies the contribution of each feature to the performance of a trained model. It is based on the principle that permuting (i.e., randomly shuffling) the values of a single feature disrupts the relationship between that feature and the target. The resulting drop in model performance indicates the importance of the feature.

We conduct feature selection on the validation set using the validation loss as the evaluation metric. In this process, only the photometric data network is utilised, as we focus exclusively on photometric data features. We first train the model using all 97 features, and then apply the PFI method by individually shuffling each feature and measuring the resulting change in validation loss. The features are then ranked by their performance impact in descending order, and feature subsets are iteratively selected from top to bottom to identify an optimal combination. Ultimately, a subset of 85 features is selected, which yields the best performance on the validation set. The selected feature patterns are listed in \ref{sec:features}.

Based on the selected set of 85 features, we further evaluate two simplified subsets: one containing only model magnitude features, and the other including model magnitude features along with adjacent-band colour features only. The results of the comparison are presented in Table~\ref{tab:feature_selection}. Both subsets result in degraded performance relative to the full 85-feature set, suggesting that the selected features are complementary and not significantly redundant.
Consequently, all model evaluations in this study are conducted using the full set of 85 features.

\begin{table}[]
    \centering
    \caption{Performance comparison across different photometric features.}
    \begin{tabular}{>{\raggedright\arraybackslash}p{4.5cm}lll}
    \hline
    Feature Set & Bias & $\sigma_{\mathrm{NMAD}}$ & $\eta$\\
    \hline
    \textbf{Optimal Features} & \textbf{0.0000} & \textbf{0.0160} & \textbf{0.0284} \\
    Only Magnitude Features & 0.0006 & 0.0280 & 0.0365 \\
    Only Magnitude and Adjacent-band Colour Features & -0.0001 & 0.0178 & 0.0313 \\
    \hline
    \end{tabular}

    \label{tab:feature_selection}
\end{table}

\subsection{List of selected photometric features}\label{sec:features}
The 85 photometric features from the DESI LS10 catalogue ranked by PFI, used as the input to the model are as follows:
$MAG\_R - MAG\_W2$,  $MAG\_I - MAG\_W1$, $MAG\_R - MAG\_W1$, $MAG\_I - MAG\_W2$, $MAG\_Z - MAG\_W1$, \\$MAG\_R - MAG\_I$, $MAG\_G  - MAG\_W2$, $MAG\_Z - MAG\_W2$, $MAG\_G - MAG\_R$, $MAG\_I - MAG\_Z$, $MAG\_G - MAG\_I$, $APMAG\_W1\_2 - APMAG\_W2\_2$, $MAG\_R - APMAG\_R\_1$, $MAG\_R - MAG\_Z$, $MAG\_G - APMAG\_G\_1$, $APMAG\_Z\_5 - APMAG\_W1\_5$, $MAG\_G$, $MAG\_G - MAG\_W1$, $MAG\_I - APMAG\_I\_1$, $APMAG\_W1\_3 - APMAG\_W2\_3$, $MAG\_I$, $MAG\_Z$, $APMAG\_I\_7 - APMAG\_Z\_7$, $MAG\_Z - APMAG\_Z\_1$, $APMAG\_R\_6 - APMAG\_I\_6$, $APMAG\_I\_6 - APMAG\_Z\_6$, $APMAG\_W1\_1 - APMAG\_W2\_1$, $MAG\_R$, $APMAG\_W2\_1 - APMAG\_W2\_2$, $APMAG\_I\_8 - APMAG\_Z\_8$, $APMAG\_R\_7 - APMAG\_I\_7$, $APMAG\_Z\_4 - APMAG\_W1\_4$, $APMAG\_G\_6 - APMAG\_R\_6$, $APMAG\_G\_7 - APMAG\_R\_7$, $APMAG\_G\_1 - APMAG\_R\_1$, $APMAG\_W1\_4 - APMAG\_W2\_4$, \\$APMAG\_Z\_1 - APMAG\_W1\_1$, $APMAG\_R\_8 - APMAG\_I\_8$, $APMAG\_I\_1 - APMAG\_I\_2$, $APMAG\_W1\_1 - APMAG\_W1\_2$, $APMAG\_W2\_3 - APMAG\_W2\_4$, $APMAG\_R\_5 - APMAG\_I\_5$, $APMAG\_W2\_2 - APMAG\_W2\_3$, $APMAG\_R\_1 - APMAG\_I\_1$, $APMAG\_G\_5 - APMAG\_R\_5$, $APMAG\_I\_2 - APMAG\_Z\_2$, $APMAG\_W2\_4 - APMAG\_W2\_5$, $APMAG\_G\_8 - APMAG\_R\_8$, $MAG\_W1 - APMAG\_W1\_1$, $APMAG\_R\_2 - APMAG\_I\_2$, $APMAG\_I\_5 - APMAG\_Z\_5$, $APMAG\_G\_1 - APMAG\_G\_2$, $MAG\_W2 - APMAG\_W2\_1$, $APMAG\_W1\_4 - APMAG\_W1\_5$, $APMAG\_R\_1 - APMAG\_R\_2$, $APMAG\_I\_1 - APMAG\_Z\_1$, $MAG\_W2$, $MAG\_G - MAG\_Z$, $APMAG\_Z\_1 - APMAG\_Z\_2$, $APMAG\_G\_4 - APMAG\_R\_4$, $APMAG\_R\_4 - APMAG\_I\_4$, $MAG\_W1$, $APMAG\_Z\_3 - APMAG\_W1\_3$, $APMAG\_I\_7 - APMAG\_I\_8$, $MAG\_W1 - MAG\_W2$, $APMAG\_Z\_2-$\\ $APMAG\_W1\_2$, $APMAG\_Z\_5 - APMAG\_Z\_6$, $APMAG\_I\_4 - APMAG\_Z\_4$, $APMAG\_G\_2 - APMAG\_R\_2$, $APMAG\_R\_3 - APMAG\_R\_4$, $APMAG\_I\_3 - APMAG\_Z\_3$, $APMAG\_Z\_7 - APMAG\_Z\_8$, $APMAG\_R\_7 - APMAG\_R\_8$, $APMAG\_G\_4 - APMAG\_G\_5$, $APMAG\_G\_3 - APMAG\_R\_3$, $APMAG\_W1\_5 - APMAG\_W2\_5$, $APMAG\_W1\_2 - APMAG\_W1\_3$, \\$APMAG\_W1\_3 - APMAG\_W1\_4$, $APMAG\_R\_5 - APMAG\_R\_6$, $APMAG\_I\_5 - APMAG\_I\_6$, $APMAG\_G\_2 - APMAG\_G\_3$, $APMAG\_Z\_3 - APMAG\_Z\_4$, $APMAG\_R\_3 - APMAG\_I\_3$, $APMAG\_Z\_6 - APMAG\_Z\_7$, $APMAG\_I\_4 - APMAG\_I\_5$.

\section{Classification and Regression in photometric redshift estimation} \label{sec:class_vs_regre}
In the redshift estimation task, two primary approaches are commonly employed: classification and regression. The distinct nature of these approaches results in different outputs from the final layer of the MLP. To assess which method yields superior performance, we conduct an experiment comparing both approaches. Both models are trained on the same dataset, and their performance metrics are computed. The results are presented in Table \ref{tab:clas_regr_res}.

\begin{table}[h]
    \centering
    \caption{Performance comparison for classification and regression approaches.}
    \begin{tabular}{cccc}
    \hline
        Model & Bias & $\sigma_{\mathrm{NMAD}}$& $\eta$  \\
        \hline
        \textbf{Classification} & \textbf{0.0002} & \textbf{0.0140} & \textbf{0.0257} \\
        Regression & -0.0025 & 0.0222 & 0.0259 \\
        \hline
    \end{tabular}
    
    \label{tab:clas_regr_res}
\end{table}

The classification approach demonstrates superior statistical performance compared to the regression method. This advantage likely arises from the inherent limitations of regression, where collapsing the prediction result into a single point estimate fails to account for the underlying uncertainty, often resulting in larger errors. Although the evaluation of the classification approach is also based on point estimates derived from the PDF, the calculation incorporates the shape of the full PDF. This allows the final point estimate to retain richer information, thereby enhancing prediction accuracy.

Moreover, the classification approach enables the direct output of a PDF, which facilitates the estimation of uncertainty -- a critical aspect of $z_\mathrm{{phot}}$ estimation. This ability to quantify uncertainty is particularly important for scientific analyses. For these reasons, we adopt the classification approach for estimating $z_\mathrm{{phot}}$ in this study.

\end{document}